\RequirePackage{lineno}
\documentclass[twocolumn,aps,tightenlinefs,superscriptaddress,preprintnumbers]{revtex4}
\usepackage{graphicx,color,dcolumn,booktabs,bm}
\usepackage{longtable,lscape}
\usepackage{amsmath}
\usepackage{indentfirst}
\usepackage{epsfig}
\usepackage{feynmf}
\usepackage{epstopdf}
\usepackage{slashed}
\usepackage{cases}
\usepackage{color}
\usepackage{multirow}
\usepackage{ulem}
\usepackage{graphicx,color,dcolumn,booktabs,bm}
\usepackage{verbatim}
\usepackage[colorlinks,linkcolor=blue,anchorcolor=green,citecolor=blue]{hyperref}
\usepackage{mathrsfs}
\usepackage{rotating}
\usepackage{threeparttable}
\usepackage{dcolumn}
\usepackage{orcidlink}

\newcommand{\lum}{{\cal L}}

\newcommand{\BR}{{\cal B}}

\newcommand{\pim}{\pi^-}

\newcommand{\EE}{e^+e^-}

\newcommand{\beq}{\begin{equation}}
\newcommand{\eeq}{\end{equation}}
\newcommand{\bitm}{\begin{itemize}}
\newcommand{\eitm}{\end{itemize}}

\begin{document}
\hyphenpenalty=10000


\title{\vspace*{0.8cm}\boldmath Observation of the decays $B^{+} \to \Sigma_{c}(2455)^{++} \bar{\Xi}_{c}^{\prime-}$ 
and $B^{0} \to \Sigma_{c}(2455)^{0} \bar{\Xi}_{c}^{\prime0}$}

\author{M.~Abumusabh\,\orcidlink{0009-0004-1031-5425}} 
\author{A.~Aggarwal\,\orcidlink{0000-0002-5623-3896}} 
\author{H.~Ahmed\,\orcidlink{0000-0003-3976-7498}} 
\author{J.~K.~Ahn\,\orcidlink{0000-0002-5795-2243}} 
\author{Y.~Ahn\,\orcidlink{0000-0001-6820-0576}} 
\author{M.~Akdag\,\orcidlink{0009-0004-3728-1077}} 
\author{N.~Akopov\,\orcidlink{0000-0002-4425-2096}} 
\author{S.~Alghamdi\,\orcidlink{0000-0001-7609-112X}} 
\author{M.~Alhakami\,\orcidlink{0000-0002-2234-8628}} 
\author{N.~Althubiti\,\orcidlink{0000-0003-1513-0409}} 
% ...
\collaboration{The Belle and Belle II Collaborations}

\begin{abstract}
We report the first observation of the decays $B^{+} \to \Sigma_{c}(2455)^{++} \bar{\Xi}_{c}^{\prime-}$ and 
$B^{0} \to \Sigma_{c}(2455)^{0} \bar{\Xi}_{c}^{\prime0}$, with significances of $6.4\,\sigma$ and $5.3\, \sigma$, 
respectively, including systematic uncertainties. 
This analysis is based on data samples containing $771.6 \times 10^{6}$ $\Upsilon(4S)$ decays
collected with the Belle detector at the KEKB collider and $520.6 \times 10^{6}$ $\Upsilon(4S)$ decays
collected with the Belle~II detector at the SuperKEKB collider.
The branching fractions are measured to be 
$\mathcal{B}(B^+ \to \Sigma_c(2455)^{++} \bar{\Xi}_c^{\prime -}) = (1.69 \pm 0.31 \pm 0.12^{+1.39}_{-0.56}) \times 10^{-3}$ 
and $\mathcal{B}(B^0 \to \Sigma_c(2455)^{0} \bar{\Xi}_c^{\prime 0}) = (1.28 \pm 0.32 \pm 0.10^{+0.26}_{-0.22}) \times 10^{-3}$,
where the first and second uncertainties are statistical and systematic, respectively, and the third arises from the uncertainties 
in the absolute branching fractions of $\bar{\Xi}_{c}^{-}$ and $\bar{\Xi}_{c}^{0}$ decays. This result represents the first 
observation of $B$-meson decays into a pair of charmed baryon-antibaryon states belonging to the same $SU(3)$ flavor sextet.	
\end{abstract}

\maketitle


$B$-meson decays provide a unique laboratory to explore the 
interplay between weak and strong interactions in the nonperturbative regime of quantum chromodynamics.
Two-body baryonic $B$ decays proceed at tree level through $W$-exchange, 
$W$-annihilation, and internal $W$-emission topologies~\cite{Cheng:2006nm, Chistov:2016kae}. 
The factorizable $W$-exchange and $W$-annihilation contributions are expected to be helicity suppressed~\cite{BaBar:2014omp}. 
However, in contrast to mesonic $B$ decays, the internal $W$-emission contribution is not necessarily color suppressed, 
and can therefore play a dominant role in baryonic two-body decays~\cite{BaBar:2014omp, Hsiao:2019wyd, Hsiao:2023mud}. 
The large mass of the $B$ meson allows for decays into charmed baryon-antibaryon pairs.
Over the past two decades, several such modes have been observed~\cite{Belle:2005gtu, BaBar:2007xtc, Belle:2019pze, LHCb:2025ueu}. 
The branching fractions of $B$-meson decays into charmed baryon–antibaryon pairs are significantly larger than those of 
two-body decays containing one charmed baryon and one charmless baryon, and both exceed those of fully charmless baryonic decays,
reaching values as large as $\mathcal{O}(10^{-3})$~\cite{ParticleDataGroup:2024cfk}.

\begin{figure}[htbp]
	\centering
	{\includegraphics[width=9cm]{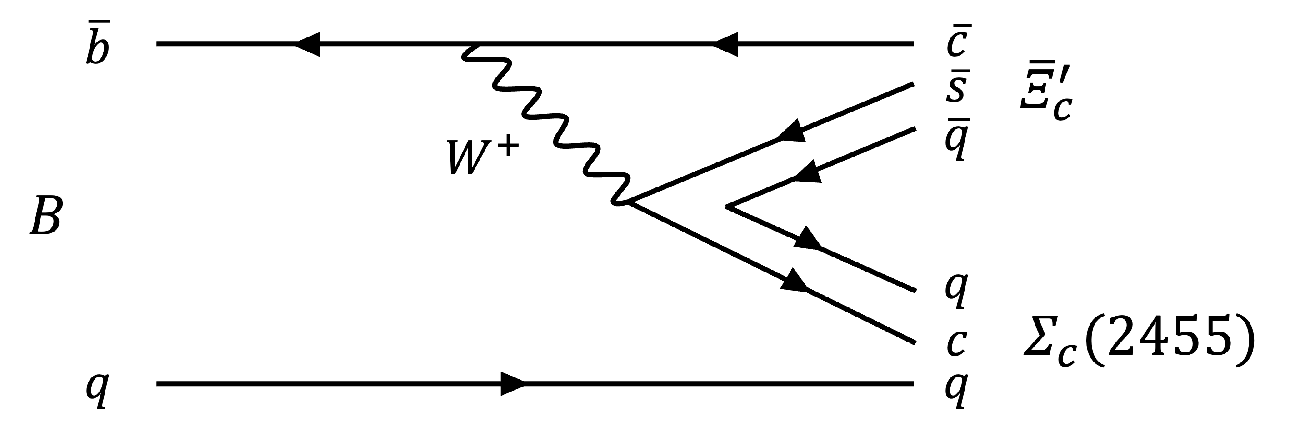}}
   \caption{Diagram illustrating the internal $W$-emission topology for the decays 
   	$B^{+} \to \Sigma_{c}(2455)^{++} \bar{\Xi}_{c}^{\prime-}$ and 
   	$B^{0} \to \Sigma_{c}(2455)^{0} \bar{\Xi}_{c}^{\prime0}$, 
   	corresponding to $q = u$ and $q = d$, respectively.}
   \label{fig1}
\end{figure}

Recently, the decays $B^{+} \to \Sigma_{c}(2455)^{++}\,\bar{\Xi}_{c}^{-}$ and 
$B^{0} \to \Sigma_{c}(2455)^{0}\,\bar{\Xi}_{c}^{0}$ were observed for the first time, by the 
Belle and Belle~II experiments~\cite{Belle:2025nup}, with branching fractions measured to be of the order of $10^{-4}$. These modes proceed through internal $W$-emission topology. 
The $\bar{\Xi}_{c}^{\prime}$ and $\bar{\Xi}_{c}$ baryons share the same valence-quark content
but in a heavy quark and light diquark model differ in the spin configuration of the light diquark.
Consequently, the decays $B^{+} \to \Sigma_{c}(2455)^{++} \bar{\Xi}_{c}^{\prime-}$ and 
$B^{0} \to \Sigma_{c}(2455)^{0} \bar{\Xi}_{c}^{\prime0}$ are expected to proceed through the same internal $W$-emission topology, as illustrated in Fig.~\ref{fig1}. 
A comparison between the branching fractions of $B \to \Sigma_{c}(2455)\,\bar{\Xi}_{c}^{\prime}$ and 
$B \to \Sigma_{c}(2455)\,\bar{\Xi}_{c}$ could provide insight into the underlying dynamics of $B$ decays to charmed baryon-antibaryon pairs belonging to different $SU(3)$ flavor multiplets. 
To date, there have been no observations of $B$ decays into charmed baryon-antibaryon
pairs in which both daughters belong to a flavor sextet.

In this paper, we report the first measurement of the decays $B^{+} \to \Sigma_{c}(2455)^{++} \bar{\Xi}_{c}^{\prime-}$ 
and $B^{0} \to \Sigma_{c}(2455)^{0} \bar{\Xi}_{c}^{\prime0}$. 
Most of the selection criteria follow those in Ref.~\cite{Belle:2025nup}.  In addition there are further
requirements related to the additional photon from the $\bar{\Xi}_{c}^{\prime}$ decay.
The $\Sigma_{c}(2455)^{++,0}$ baryons are reconstructed via their decays into $\Lambda_{c}^{+}\pi^{\pm}$,
with the $\Lambda_{c}^{+}$ reconstructed using the $\Lambda_{c}^{+} \to p K^{-} \pi^{+}$ and
$\Lambda_{c}^{+} \to p K_{S}^{0}$ modes.
The $\bar{\Xi}_{c}^{\prime-,0}$ baryons are reconstructed via their decays into $\gamma\bar{\Xi}_{c}^{-,0}$,
with the $\bar{\Xi}_{c}^{-}$ reconstructed using the $\bar{\Xi}_{c}^{-} \to \bar{\Xi}^{+} \pi^{-} \pi^{-}$ and
$\bar{\Xi}_{c}^{-} \to \bar{p}K^{+}\pi^{-}$ modes and the $\bar{\Xi}_{c}^{0}$ reconstructed using the
$\bar{\Xi}_{c}^{0} \to \bar{\Xi}^{+} \pi^{-}$ and $\bar{\Xi}_{c}^{0} \to \bar{\Lambda} K^{+} \pi^{-}$ modes.
Charge-conjugate channels are implicitly included throughout this paper. 
This study is based on data samples containing $(771.6 \pm 10.6) \times 10^{6}$ 
and $(520.6 \pm 7.6) \times 10^{6}$ $\Upsilon(4S)$ events collected with the 
Belle~\cite{Belle:2000cnh} and Belle~II~\cite{Belle-II:2010dht} detectors, 
respectively, at the $e^{+}e^{-}$ center-of-mass energy ($\sqrt{s}$) of 
10.58\,GeV, corresponding to integrated luminosities of 
711\,$\mathrm{fb}^{-1}$~\cite{Belle:2012iwr} and 
492\,$\mathrm{fb}^{-1}$~\cite{Belle-II:2024vuc}.
The analysis strategy is to fully reconstruct the signal decay chains and extract the
signal yields from simultaneous fits to
the distributions of the difference between the reconstructed $B$-meson energy and the beam energy at 
the $\Upsilon(4S)$ rest frame, both in the $\bar{\Xi}_c^{\prime}$  signal and sideband regions.

The Belle detector was a cylindrical large-solid-angle magnetic spectrometer located at the interaction point (IP) of the KEKB 
asymmetric-energy $e^+e^-$ collider~\cite{Kurokawa:2001nw}. The detector consisted of a silicon vertex detector (SVD), a central 
drift chamber (CDC), an array of aerogel threshold Cherenkov counters (ACC), a barrel-like arrangement of time-of-flight (TOF)
scintillation counters, and an electromagnetic calorimeter (ECL) composed of CsI(Tl) crystals located inside a superconducting solenoid 
coil that provided a $1.5~\hbox{T}$ axial magnetic field. An iron flux-return located outside the coil was instrumented to detect 
$K^{0}_{L}$-mesons and to identify muons. A detailed description of the Belle detector can be found in Ref.~\cite{Belle:2000cnh}.

The Belle II detector, located at the SuperKEKB asymmetric-energy $\EE$ collider~\cite{Akai:2018mbz}, is an upgraded
version of the Belle detector. Belle~II includes a SVD consisting of pixel sensors and double-sided strip detectors, 
and a CDC. The CDC is surrounded by two types of Cherenkov light detector systems: time-of-propagation (TOP)
detector for the barrel region and an aerogel ring-imaging Cherenkov (ARICH) detector for the forward end-cap region. 
The Belle ECL is reused in Belle II along with the solenoid and the iron flux-return yoke. However, the ECL readout system 
has been upgraded. The solenoid flux return is instrumented with resistive-plate chambers
and plastic scintillator modules to detect muons and $K^{0}_{L}$ mesons.
A detailed description of the Belle~II detector can be found in Ref.~\cite{Belle-II:2010dht}.

Simulated signal events are used to optimize the selection criteria, 
calculate the reconstruction efficiencies, and determine the fit models. 
The \textsc{evtgen}~\cite{Lange:2001uf} and \textsc{pythia}~\cite{Sjostrand:2006za,Sjostrand:2014zea} 
software packages are used to generate $e^+e^- \to \Upsilon(4S) \to B \bar{B}$ events
with final-state radiation simulated by the \textsc{photos} software package~\cite{Barberio:1990ms}. In the simulation, 
one $B$ meson is not constrained to decay into a specific final state, while the other decays into a signal mode.
The simulation includes beam-induced background data overlay~\cite{Liptak:2021tog}.
Inclusive simulated samples of $e^{+}e^{-} \to q\bar{q}$ ($q = u, d, s, c$) and $\Upsilon(4S) \to B\bar{B}$
are used to optimize the selection criteria and identify the background sources~\cite{Zhou:2020ksj},
corresponding to six times the integrated luminosity of the Belle data and four times that of the Belle~II data.
The \textsc{kkmc}~\cite{Jadach:1999vf}  and \textsc{pythia}~\cite{Sjostrand:2006za,Sjostrand:2014zea} 
software packages are used to simulate the $e^{+} e^{-} \to q\bar{q}$ processes.  
The detector responses are modeled by the software packages \textsc{geant3}~\cite{GEANT3} 
for Belle and \textsc{geant4}~\cite{GEANT4:2002zbu} for Belle II.

We analyze both Belle and Belle~II data using the Belle~II analysis software framework~\cite{Kuhr:2018lps}. 
For the Belle data, the event reconstruction up to the particle level is performed with the Belle software, 
and the reconstructed data are then converted into a Belle~II compatible format~\cite{Gelb:2018agf}. 
This allows the Belle and Belle~II samples to be analyzed within a unified framework.
To avoid experimental bias, the signal region was not examined until the analysis procedure was finalized.
All selection criteria are determined by iteratively optimizing the Punzi figure-of-merit for a target significance of five standard deviations in simulation~\cite{Punzi}.
We apply identical event-selection criteria in the Belle and Belle~II analyses, unless otherwise stated.

We select tracks, except for those from the decays of $K_S^0$, $\bar{\Lambda}$, and $\bar{\Xi}^+$, 
by requiring the impact parameters relative to the $e^{+}e^{-}$ IP to be less than 2.0\,cm perpendicular 
to the $z$-axis and less than 4.0\,cm parallel to it.
The $z$-axis is defined as the central solenoid axis with the positive direction 
in the $e^-$ beam direction, common to both Belle and Belle~II.
The identification of charged tracks uses the likelihood ratio 
$\mathcal{R}(h|h^{\prime}) = \mathcal{L}(h)/[\mathcal{L}(h) + \mathcal{L}(h^{\prime})]$, where $\mathcal{L}(h^{(\prime)})$ 
is the likelihood of the charged track being a hadron $h^{(\prime)} = p$, $K$, or $\pi$.
This likelihood ratio is determined with a particle identification (PID) algorithm that combines information 
from different detector subsystems, including the specific ionization in the CDC, time-of-flight measurement in the TOF (TOP), 
and the response of the ACC (ARICH) in Belle (Belle~II)~\cite{Nakano:2002jw, Belle-II:2025tpe}.
Correction factors derived from control samples in data are applied to account for differences between data and simulation
in tracking and PID efficiencies.
Tracks with $\mathcal{R}(p|K) > 0.6$ and $\mathcal{R}(p|\pi) > 0.6$ are identified as proton candidates; 
charged kaon (pion) candidates must satisfy $\mathcal{R}(K|\pi) > 0.6$ ($<0.4$). 
The efficiencies of these requirements range from 80\% to 95\%, with corresponding 
misidentification rates between 3\% and 9\%. The pion candidates used to reconstruct the 
$K_{S}^{0}$, $\bar{\Lambda}$, and $\bar{\Xi}^{+}$ candidates are exempt from PID requirements as their
kinematics provide sufficient discrimination.

The $K_{S}^{0}$ candidates are reconstructed from pairs of oppositely charged tracks 
treated as pions. An artificial neural network~\cite{Belle:2018xst} and a boosted decision tree (BDT) 
classifier~\cite{Belle:2021efh} are used in Belle and Belle~II,
respectively, to select the $K_{S}^{0}$ candidates.
The $\bar{\Lambda}$ candidates are reconstructed from the decay $\bar{\Lambda} \to \bar{p}\pi^{+}$ 
and selected using momentum-dependent requirements on variables describing the displaced decay topology in Belle~\cite{Belle:2020xku}
and using a BDT classifier in Belle~II.
These discriminators primarily rely on the kinematic properties of the 
$K_{S}^{0}$ or $\bar{\Lambda}$ and their decay products. 
The invariant masses of the $K_{S}^{0}$ and $\bar{\Lambda}$ candidates 
are required to be within $\pm 9.0$\,MeV/$c^{2}$ and $\pm 5.5$\,MeV/$c^{2}$ 
of their known masses~\cite{ParticleDataGroup:2024cfk}, respectively, 
corresponding to approximately 2.5 times the mass resolution ($\sigma$).
The selected $\bar{\Lambda}$ candidate is then combined with a $\pi^{+}$ candidate to form a $\bar{\Xi}^+$ candidate. 
The invariant mass of $\bar{\Xi}^+$ candidates is required to be within $\pm 6.5$\,MeV/$c^2$ of
its known mass~\cite{ParticleDataGroup:2024cfk} (approximately $2.5\, \sigma$).
A vertex fit is performed for each $K_{S}^{0}$, $\bar{\Lambda}$, and $\bar{\Xi}^{+}$ candidate, 
and the candidate mass is constrained to the corresponding known value~\cite{ParticleDataGroup:2024cfk}.
	
We define a photon candidate as an energy deposit in the ECL not associated with
the extrapolated trajectory of any charged track. To suppress backgrounds from neutral hadrons, we require
$E(3\times3)/E(5\times5) \geq 0.6$, where $E(n\times n)$ denotes the energy deposited 
in an $n\times n$ array of crystals centered on the highest-energy crystal (for Belle~II,  the four corner crystals of  the $5\times5$ array are excluded).  
The photon energy is required to exceed 50\,MeV in the laboratory frame.
At Belle~II, BDT classifiers are used to suppress energy deposits from secondary hadronic interactions unrelated 
to the relevant collision and overlapping beam-induced photons~\cite{Cheema:2024iek}.
The combined signal efficiencies (background rejection rates) of the two BDTs are approximately
89\% (47\%) and 89\% (44\%) for the charged and neutral channels, respectively.

The invariant masses of the reconstructed $\Lambda_c^+$, $\bar{\Xi}_c^-$, and $\bar{\Xi}_c^0$ 
candidates are required to lie within $\pm 15$, $\pm 18$, and $\pm 18\,{\rm MeV}/c^2$ of their known 
values~\cite{ParticleDataGroup:2024cfk}, respectively, corresponding in each case to approximately $2.5\, \sigma$.
Each $\Lambda_{c}^{+}$ and $\bar{\Xi}_{c}^{-,0}$ candidate is fitted to a common vertex, and the candidate mass is constrained to the corresponding known value~\cite{ParticleDataGroup:2024cfk}.
The $\Sigma_{c}(2455)^{++,0}$ baryons are reconstructed via their decays into
$\Lambda_{c}^{+}\pi^{\pm}$, with the signal region defined as 
$2.446 < M(\Lambda_{c}^{+}\pi^{\pm}) < 2.464\,\mathrm{GeV}/c^{2}$. 
After applying this requirement, more than 95\% of the signal events are retained.
The sideband regions of $M(\Lambda_{c}^{+}\pi^{\pm})$ are defined as 
$2.436 < M(\Lambda_{c}^{+}\pi^{\pm}) < 2.444\,\mathrm{GeV}/c^{2}$ and 
$2.466 < M(\Lambda_{c}^{+}\pi^{\pm}) < 2.494\,\mathrm{GeV}/c^{2}$.
The $\bar{\Xi}_{c}^{\prime -,0}$ baryons are reconstructed via their  decays into $\gamma \bar{\Xi}_{c}^{-,0}$, 
with the reconstructed masses required to be within $\pm 20$\,MeV/$c^{2}$ of the known 
$\bar{\Xi}_{c}^{\prime -,0}$ masses~\cite{ParticleDataGroup:2024cfk}, corresponding in each case to approximately $2.5\, \sigma$.
Vertex fits are applied to the $\Sigma_{c}(2455)^{++,0}$ and $\bar{\Xi}_{c}^{\prime -,0}$ candidates, 
with their masses constrained to the known values~\cite{ParticleDataGroup:2024cfk}.

The $B^{+}$ and $B^{0}$ candidates are reconstructed from the 
$\Sigma_{c}(2455)^{++} \bar{\Xi}_{c}^{\prime-}$ and $\Sigma_{c}(2455)^{0} \bar{\Xi}_{c}^{\prime0}$ 	combinations, respectively.
This implies four distinct reconstruction modes for each signal channel. 
A vertex fit is applied to the $B$ candidates with a requirement of 
$\chi^{2}/{\rm ndf} < 20$, where ndf is the number of degrees of freedom.
The beam-constrained mass $M_{\rm bc}$ and the energy difference $\Delta E$ 
are defined for $B$ candidates as $M_{\rm bc} = \sqrt{E_{\rm beam}^2 - \left(\sum_i \vec{p}_i\right)^2}$
and $\Delta E = \sum_i E_i - E_{\rm beam}$, where $E_{\rm beam} = \sqrt{s}/2$ 
is the beam energy  and ($E_{i}$, $\vec{p}_i$) is the four-momentum 
of the $i$th daughter of the $B$ meson, all calculated in the $e^{+}e^{-}$ c.m.\ system.
Candidates satisfying $M_{\rm bc} > 5.27\,{\rm GeV}/c^2$ (approximately  $2.5\, \sigma$)
and $|\Delta E| < 0.06\,{\rm GeV}$ (approximately  $12\, \sigma$) are retained for further analysis.
The scatter plots of $M(\Lambda_c^+\pi^\pm)$ versus
$M(\gamma\bar{\Xi}_c^{-,0})$ for reconstructed $B^{+}\to\Sigma_c(2455)^{++}\bar{\Xi}_c^{\prime-}$ and
$B^{0}\to\Sigma_c(2455)^{0}\bar{\Xi}_c^{\prime0}$ candidates in the signal and sideband regions of $\Delta E$ from the combined Belle and Belle~II data samples
are shown in Appendix~\ref{App_A}, along with their one-dimensional projections onto 
$M(\Lambda_c^+\pi^\pm)$ and $M(\gamma\bar{\Xi}_c^{-,0})$ in the  signal region of $\Delta E$.
Because the photon from the $\bar{\Xi}_c^{\prime}$ decay has very low energy, 
many random low-energy photon candidates, such as residual neutral clusters in the electromagnetic calorimeter 
or photons from other particles in the event, can form incorrect combinations with the $\bar{\Xi}_{c}$ candidates. 
This results in a large number of events with multiple candidates.
After applying all the selection criteria, the fractions of events in signal simulation with multiple candidates
are $60.5\%$--$63.8\%$ ($54.3\%$--$58.9\%$) for the charged channel and $61.0\%$--$64.0\%$ ($53.8\%$--$58.4\%$) 
for the neutral channel in Belle (Belle~II),  depending on the reconstruction mode. 
The corresponding average candidate multiplicities are
2.12--2.33 (1.93--2.13) for the charged channel and 2.15--2.29 (1.93--2.03) for the neutral channel in Belle (Belle~II).
For events with multiple candidates, a best-candidate selection (BCS) is applied, 
retaining the candidate with the smallest total $\chi^{2}$  from the vertex and mass-constrained 
fits to the charmed baryons and the $B$ meson. 
It was checked that the BCS procedure does not introduce peaking structures in the $\Delta E$ distribution or distort its shape, 
and yields signal efficiencies of $73.0\%$--$76.2\%$ ($72.5\%$--$75.1\%$) for the charged channel and
$73.5\%$--$74.0\%$ ($72.0\%$--$75.0\%$) for the neutral channel in Belle (Belle~II).

It is estimated from simulation that after the BCS around 50\% of the selected
signal candidates are incorrectly reconstructed, and these are referred to as self-cross-feed (SCF) events.
These SCF events mainly originate from the reconstruction of the
$\bar{\Xi}_c^{\prime-,0}\to\gamma\bar{\Xi}_c^{-,0}$ decays, in which the genuine low-energy photon has been replaced by a random low-energy photon candidate.
The SCF component exhibits a broad $\Delta E$ distribution that is clearly distinguishable from the correctly reconstructed signal but overlaps significantly 
with the combinatorial background. The SCF events are not treated as signal when determining the branching fractions. 
To have better control of SCF, we define the sideband regions of $M(\gamma\bar{\Xi}_c^{-,0})$ as 
$2.508 < M(\gamma\bar{\Xi}_c^{-,0}) < 2.548\,\mathrm{GeV}/c^{2}$
and $2.608 < M(\gamma\bar{\Xi}_c^{-,0}) < 2.658\,\mathrm{GeV}/c^{2}$.
The centers of the sideband regions are about $5\, \sigma$ away from the
center of the signal region.
The sideband regions contain a significant SCF contribution, 
with a shape in the $\Delta E$ distribution similar to that observed in the signal region.
To better constrain the SCF component in the signal extraction, 
we perform a simultaneous fit to the $\Delta E$ distributions in the signal
and sideband regions of $M(\gamma\bar{\Xi}_c^{-,0})$.
Backgrounds are studied using inclusive simulated samples~\cite{Zhou:2020ksj} and data 
from the sidebands of $M(\Lambda_c^{+})$, $M(\Lambda_c^+\pi^{\pm})$, $M(\bar{\Xi}_{c}^{-,0})$, 
and $M_{\rm bc}$ distributions. No peaking background is observed  in the $\Delta E$ distributions of events from these sidebands.

To extract the signal yield of $B^{+} \to \Sigma_{c}(2455)^{++} \bar{\Xi}_{c}^{\prime-}$ 
or $B^{0} \to \Sigma_{c}(2455)^{0} \bar{\Xi}_{c}^{\prime0}$, 
we perform simultaneous unbinned  extended maximum-likelihood fits to the $\Delta E$ distributions
of four samples: events from the signal and sideband regions of $M(\gamma\bar{\Xi}_c^{-,0})$ in both Belle and Belle~II data.
The likelihood functions used to model the $\Delta E$ distributions
in the signal and sideband regions of  $M(\gamma\bar{\Xi}_{c}^{-,0})$  are parametrized as

\begin{displaymath}\label{fiting_function1}
	\begin{aligned}
		\mathcal{L} =  & \frac{e^{-(N_{\rm sig} + N_{\rm scf} + N_{\rm bkg})}}{N!}  \prod_{i}^{N}  [N_{\rm sig} \mathcal{F}_{\rm sig}(\Delta E_{i})  \\  
		+ &  N_{\rm scf} \mathcal{F}_{\rm scf} (\Delta E_{i})  + N_{\rm bkg} \mathcal{F}_{\rm bkg} (\Delta E_{i})]
	\end{aligned}
\end{displaymath}
and 
\begin{displaymath}\label{fiting_function2}
	\begin{aligned}
		\mathcal{L}^{\rm sb} = &  \frac{e^{-(N_{\rm sig}^{\rm sb} + N_{\rm scf}^{\rm sb} + N_{\rm bkg}^{\rm sb})}}{N^{\rm sb}!}  \prod_{i}^{N^{\rm sb}}  [N_{\rm sig}^{\rm sb} \mathcal{F}_{\rm sig}(\Delta E_{i})  \\ + &  N_{\rm scf}^{\rm sb} \mathcal{F}_{\rm scf}^{\rm sb}(\Delta E_{i})  + N_{\rm bkg}^{\rm sb} \mathcal{F}_{\rm bkg}^{\rm sb}(\Delta E_{i})],
	\end{aligned}
\end{displaymath}
respectively. Here, $\mathcal{F}^{(\rm sb)}(\Delta E_i)$ denotes the probability density function (PDF) evaluated at the measured $\Delta E$ value of the $i$-th event,
and $N^{(\rm sb)}$ denotes the number of observed events.
The signal PDF, $\mathcal{F}_{\rm sig}$, is parameterized by a double-Gaussian function with the same mean values.
All parameters of the double-Gaussian function are fixed to the values obtained from simulation.
The SCF PDFs, $\mathcal{F}_{\rm scf}^{(\rm sb)}$, are modeled using the nonparametric kernel density estimation~\cite{Cranmer:2000du} 
derived from simulation. The combinatorial background PDFs, $\mathcal{F}_{\rm bkg}^{(\rm sb)}$, 
are described by first-order polynomials with free parameters. 
The polynomial parameters are allowed to vary independently between the signal and sideband regions of $M(\gamma \bar{\Xi}_{c}^{-,0})$ and between the Belle and Belle~II data samples.
The quantities $N_{\rm sig}^{(\rm sb)}$, $N_{\rm scf}^{(\rm sb)}$, 
and $N_{\rm bkg}^{(\rm sb)}$ represent the signal, SCF, and combinatorial background yields, respectively. The signal (SCF) yields satisfy
$N_{\rm sig(scf)}^{\rm sb} = f_{\rm sig(scf)} N_{\rm sig(sc f)}$, where $f_{\rm sig(scf)}$ is defined as the ratio of the signal (SCF) yield in the sideband 
regions to that in the signal region of $M(\gamma\bar{\Xi}_c^{-,0})$, as estimated from simulation.
The values of $f_{\rm sig}$ and $f_{\rm scf}$ are fixed by simulation
to 0.02 (0.05) and 1.9 (1.8)  for the charged channel  and 0.01 (0.04) and 1.8 (1.6) for the neutral channel in Belle (Belle~II), respectively.
The event yields are determined from the fit, with the signal yields in the Belle and Belle~II 
datasets constrained to follow the expected ratio for a common branching fraction.

\begin{figure*}[htbp]
	\begin{center}
		\hspace{-0.7cm}
		\includegraphics[width=7cm]{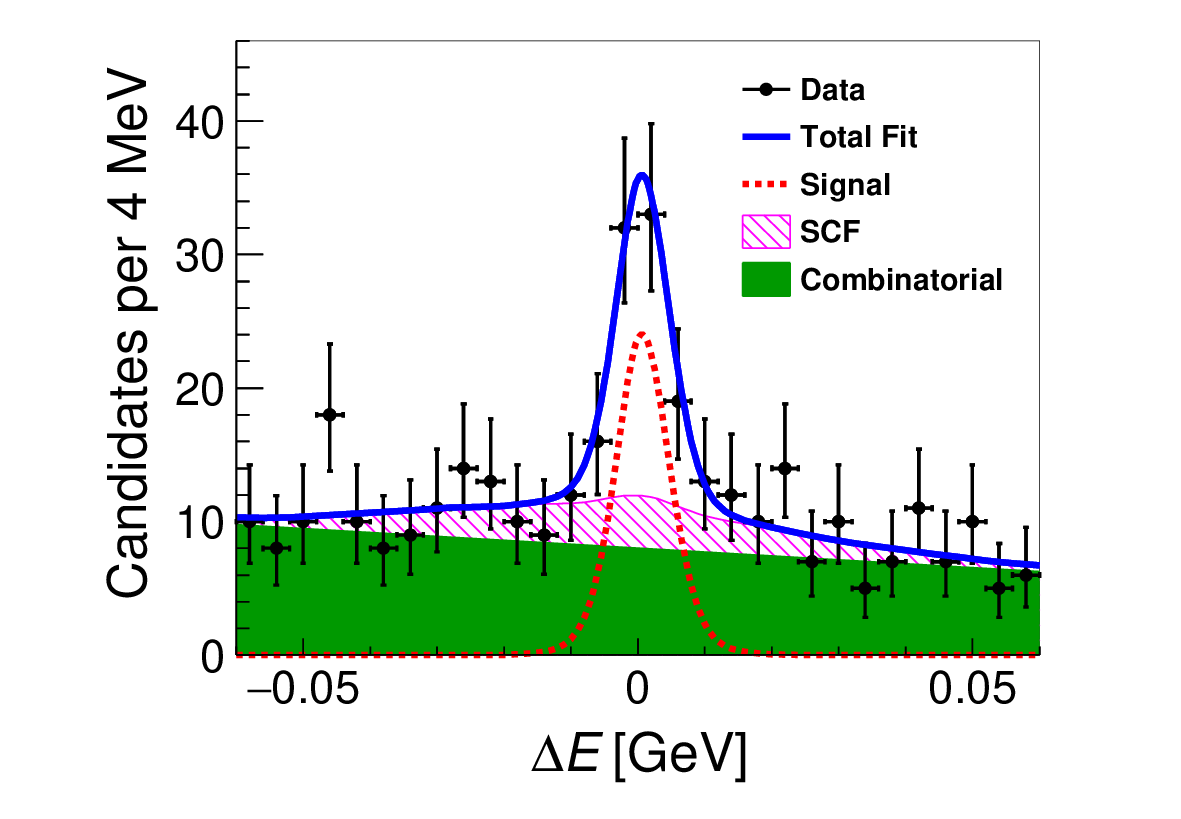}\hspace{-0.3cm}
		\includegraphics[width=7cm]{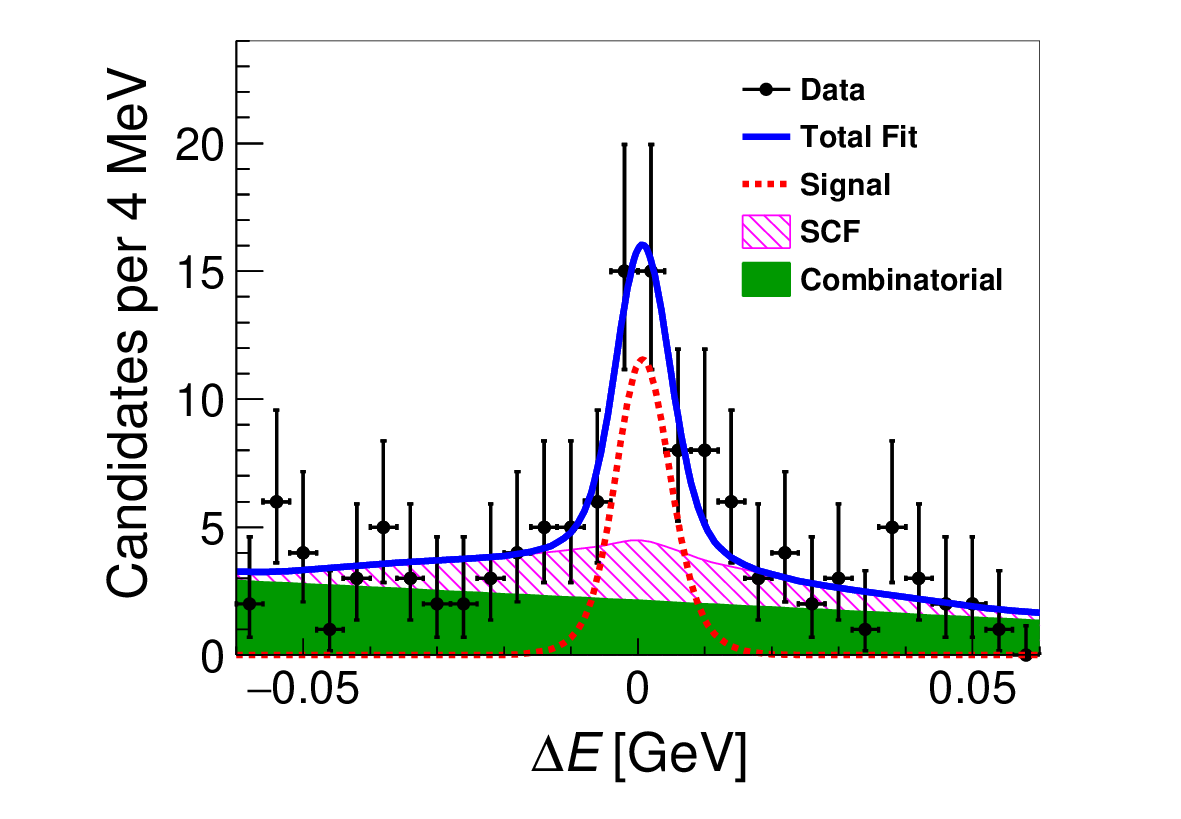}
	    \put(-290,140){$\int \lum dt$ = 711\,fb$^{-1}$ (Belle) + 492\,fb$^{-1}$ (Belle~II)}
	
		\hspace{-0.7cm}
		\includegraphics[width=7cm]{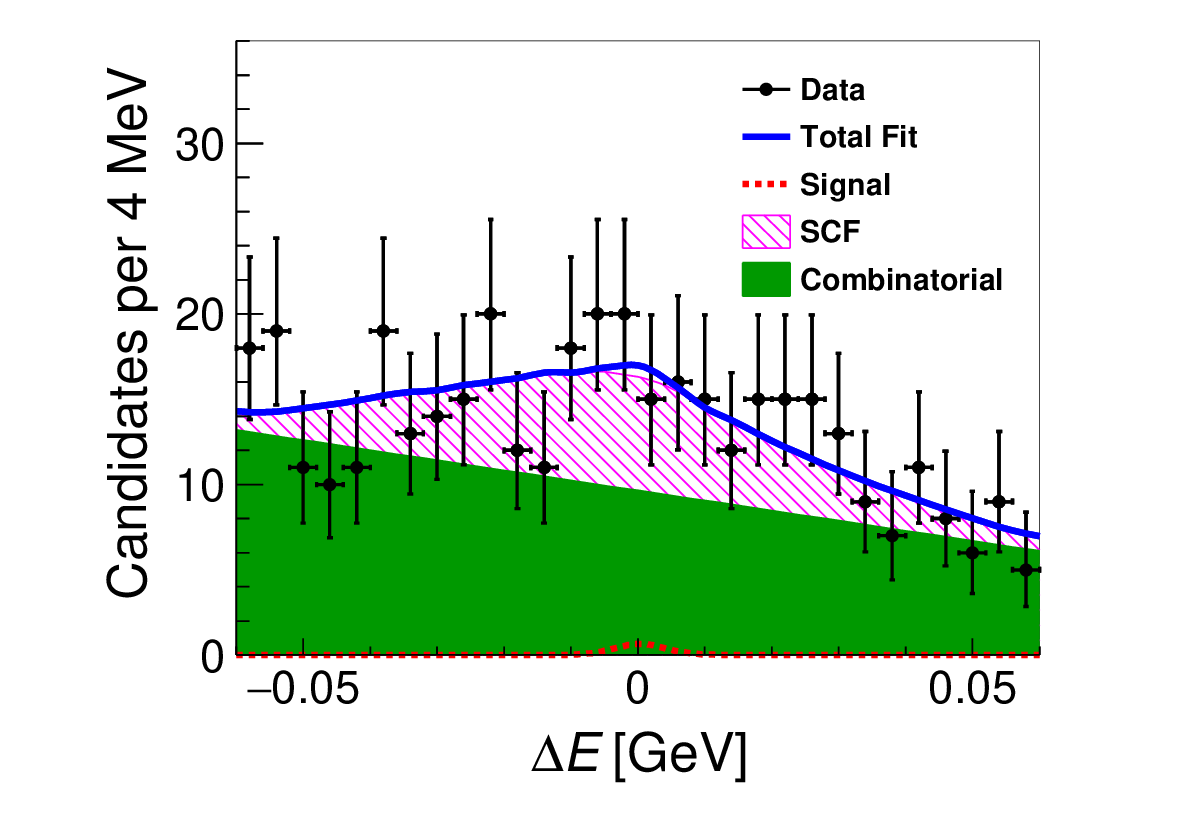}\hspace{-0.3cm}
        \includegraphics[width=7cm]{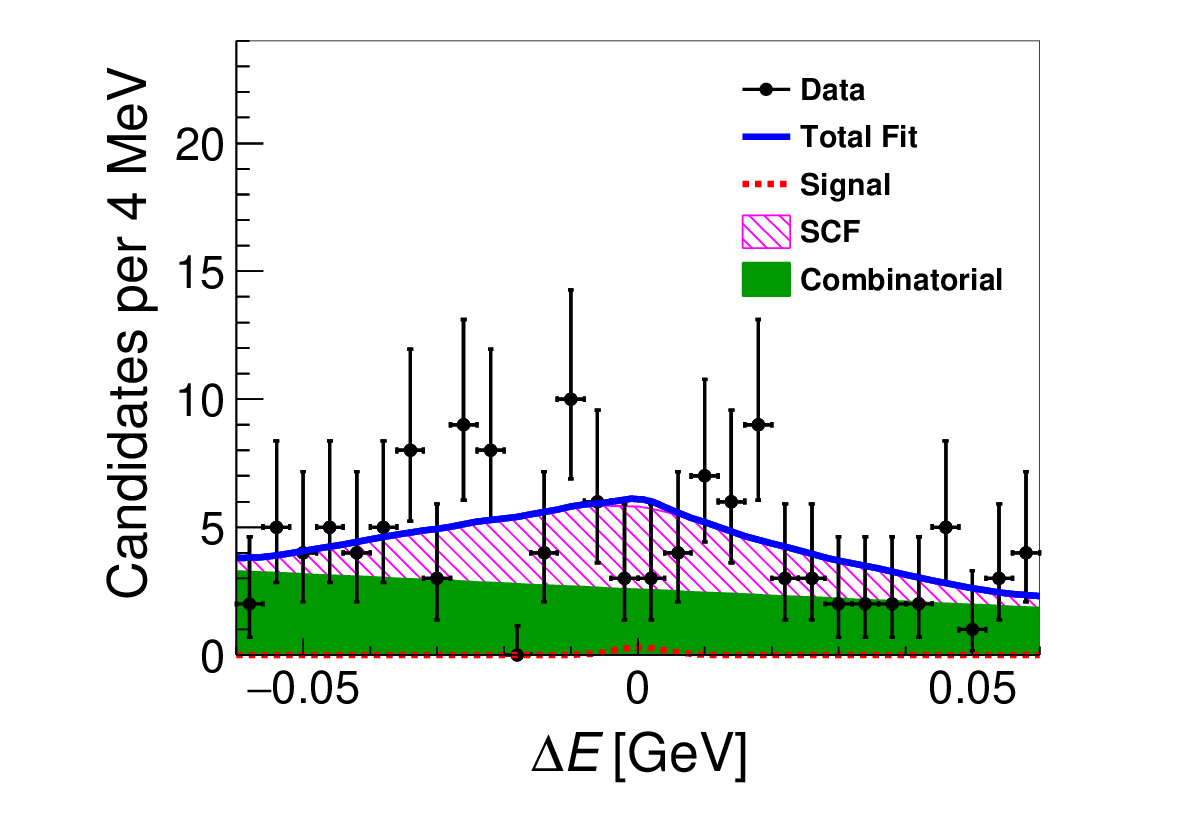}
       \caption{Distributions of $\Delta E$ for the reconstructed 
	  (left) $B^{+} \to \Sigma_{c}(2455)^{++} \bar{\Xi}_{c}^{\prime-}$ and 
	  (right) $B^{0} \to \Sigma_{c}(2455)^{0} \bar{\Xi}_{c}^{\prime0}$ candidates 
	  in the (top) signal and (bottom) sideband regions of $M(\gamma \bar{\Xi}_{c}^{-,0})$ 
	  in the combined Belle and Belle~II data. Points with error bars denote the data. 
	  The blue solid and red dashed curves represent the total fit results and fitted 
	  signal components, respectively, while the magenta hatched and green filled histograms represent the
	  fitted SCF and combinatorial background components.}\label{fig2}
	\end{center}
\end{figure*}

Figure~\ref{fig2} shows the $\Delta E$ distributions of candidates reconstructed in the signal and sideband regions of 
$M(\gamma \bar{\Xi}_{c}^{-,0})$ in the combined Belle and Belle~II data, with  fit results overlaid. The fitted signal yields for the decays 
$B^+ \to \Sigma_{c}(2455)^{++} \bar{\Xi}_{c}^{\prime-}$ and $B^{0} \to \Sigma_{c}(2455)^{0} \bar{\Xi}_{c}^{\prime0}$ 
are $62.3 \pm 11.3$ and $30.7 \pm 7.6$,  respectively, with statistical significances of $8.6\, \sigma$ and $6.9\, \sigma$.
The statistical significances are calculated using $-2\ln(\mathcal{L}_{0}/\mathcal{L}_{\rm max})$, 
accounting for the difference in the number of degrees of freedom ($\Delta\textrm{ndf=3}$),
where $\mathcal{L}_{0}$ and $\mathcal{L}_{\rm max}$ are the likelihood values obtained from fits 
without and with the signal and SCF components, respectively. 

To evaluate the signal significances including systematic uncertainties, we repeat the significance calculation using the same alternative fits 
as those used to evaluate the systematic uncertainties associated with the fit models, with the relevant variations considered in combination. 
The most conservative values, corresponding to the smallest significances obtained among all tested cases, are	
$6.4\, \sigma$ for the $B^+ \to \Sigma_{c}(2455)^{++} \bar{\Xi}_{c}^{\prime-}$ channel and 
$5.3\, \sigma$ for the $B^{0} \to \Sigma_{c}(2455)^{0} \bar{\Xi}_{c}^{\prime0}$ channel.
These values are taken as the final signal significances
after incorporating systematic effects.

The branching fractions of  $B^{+} \to \Sigma_{c}(2455)^{++} \bar{\Xi}_{c}^{\prime-}$ and
$B^{0} \to \Sigma_{c}(2455)^{0} \bar{\Xi}_{c}^{\prime0}$ decays are calculated using 
\begin{equation*}
	    \BR = \frac{N_{\rm sig}}{ 2  f_{+-/00} [N_{\Upsilon(4S)}^{\rm B} \sum_{i}(\varepsilon_{i}^{\rm B} \BR_{i}) + N_{\Upsilon(4S)}^{\rm B2} 
		\sum_{i}(\varepsilon_{i}^{\rm B2} \BR_{i})]}.
\end{equation*}
Here, $N_{\rm sig}$ represents the number of fitted $B^{+} \to \Sigma_{c}(2455)^{++} \bar{\Xi}_{c}^{\prime-}$ or
$B^{0} \to \Sigma_{c}(2455)^{0} \bar{\Xi}_{c}^{\prime0}$ signal events in the combined Belle and Belle~II datasets; 
$N_{\Upsilon(4S)}^{\rm B, B2}$ denotes the total number of $\Upsilon(4S)$ events in the Belle or Belle~II datasets;
$f_{+-/00}$ refers to the fraction of charged ($f_{+-} = 0.5113_{-0.0108}^{+0.0073}$) or neutral ($f_{00} = 0.4861_{-0.0080}^{+0.0074}$) $B\bar{B}$ pairs~\cite{HeavyFlavorAveragingGroupHFLAV:2024ctg};
the term $\sum_{i}(\varepsilon_{i}^{\rm B, B2} \BR_i)$ represents the sum of the 
reconstruction efficiencies $\varepsilon_{i}^{\rm B, B2}$ (for Belle or Belle~II) times the
corresponding secondary branching fractions $\mathcal{B}_i$ over all the reconstruction modes.
The values of $\varepsilon_i^{\rm B,B2}$ and $\varepsilon_i^{\rm B,B2}\mathcal{B}_i$ 
for the individual reconstruction modes are listed in Appendix~\ref{App_B}.
The values of the aforementioned quantities, as obtained from the fit, and 
the calculated branching fractions are summarized in Table~\ref{table1}.

\begin{table*}[htbp]
	\centering
	\caption{\label{table1} Summary of the fitted signal yields, the values of $\sum_{i}(\varepsilon_{i}^{\rm B, B2} \BR_i)$, and the measured branching fractions of $B^{+} \to \Sigma_{c}(2455)^{++} \bar{\Xi}_{c}^{\prime-}$ and $B^{0} \to \Sigma_{c}(2455)^{0} \bar{\Xi}_{c}^{\prime0}$ decays. 
	We list only the statistical uncertainties of the signal yields. For the branching fractions, the first and second uncertainties are statistical and systematic, respectively, while the third originates 
	from the absolute branching fractions of the $\bar{\Xi}_{c}^{-, 0}$ decays~\cite{ParticleDataGroup:2024cfk}.}
	\renewcommand{\arraystretch}{1.3}
	\begin{tabular}{lcccc}
		\hline\hline
		& $N_{\rm sig}$ &  $\sum_{i}(\varepsilon_i^{\rm B} \BR_i)~(10^{-5})$
		& $\sum_{i}(\varepsilon_i^{\rm B2} \BR_i)~(10^{-5})$ & $\BR~(10^{-3})$  \\
		\hline  
		$B^{+}$ &  $62.3\pm11.3$     & 2.43   & 3.32  & $1.69 \pm 0.31 \pm 0.12^{+1.39}_{-0.56}$\\
		$B^{0}$ &  $30.7\pm7.6$      & 1.53   & 2.49 & $1.28 \pm 0.32 \pm 0.10^{+0.26}_{-0.22}$\\
		\hline\hline 
	\end{tabular}
\end{table*}

The following sources of systematic uncertainties are the most important ones, and  have been considered in this work:
the detection-efficiency (DE) uncertainties ($\sigma_{\rm DE}$), 
the statistical uncertainty on the efficiency determined from simulation ($\sigma_{\rm eff}$), 
the uncertainties of the branching fractions of intermediate states ($\sigma_{\BR_{i}}$), 
the uncertainty of the total number of $\Upsilon(4S)$ events ($\sigma_{N_{\Upsilon(4S)}}$), 
the uncertainty on the fraction of charged or neutral $B\bar{B}$ events ($\sigma_{f_{+-/00}}$), 
the uncertainty associated with the BCS strategy ($\sigma_{\rm BCS}$), 
and uncertainties related to the fit models ($\sigma_{\rm fit}$). 
We explain the sources of systematics in the following, and summarize their values in Table~\ref{table2}, where the total
uncertainty ($\sigma_\mathrm{total}$) is calculated by adding the individual contributions in quadrature.

The DE uncertainties include contributions from tracking efficiency, 
PID efficiency, and the reconstruction of photon, $K_{S}^{0}$, 
and $\bar{\Lambda}$ candidates, which are estimated using data control samples. 
The tracking efficiency uncertainty depends on the particle charge, momentum, and polar angle, 
and ranges from 0.32\% to 1.22\% (0.38\% to 1.43\%) per track in Belle (Belle~II).
These uncertainties are determined from control samples of $D^{*+} \to D^{0}(\to K_S^0 \pi^+ \pi^-)\pi^{+}$ for Belle, 
and $\bar{B}^{0} \to D^{*+}(\to D^{0}\pi^+)\pi^{-}$ and $e^+e^- \to \tau^{+}\tau^{-}$ for Belle~II.
The slightly larger tracking uncertainties in Belle II result from the limited statistics of the control samples.
The PID efficiency uncertainties are estimated to be 1.0\% (0.8\%) for pions, 
1.3\% (1.1\%) for kaons, and 2.2\% (0.7\%) for protons in Belle (Belle~II)~\cite{Nakano:2002jw, Belle-II:2025tpe}. 
The photon reconstruction uncertainty is 2.0\% (1.6\%) in Belle (Belle~II), 
as determined from control samples of radiative Bhabha (radiative muon-pair) events.
The $K_{S}^{0}$ reconstruction uncertainty is 1.6\% and 2.1\% in Belle and Belle~II, respectively, 
evaluated using control samples of 
$D^{*+} \to D^{0}(\to K_{S}^{0}\pi^{0})\pi^{+}$ and 
$D^{*+} \to D^{0}(\to K_{S}^{0}\pi^{+}\pi^{-})\pi^{+}$. 
The $\bar{\Lambda}$ reconstruction uncertainty is 2.2\% and 1.3\% in Belle and Belle~II, respectively, 
evaluated using control samples of 
$\bar{\Lambda} \to \bar{p} \pi^{+}$ and 
$\bar{\Lambda}_{c}^{-} \to \bar{\Lambda}( \to \bar{p} \pi^{+})\pim$.
The individual uncertainties for the different reconstruction modes in Belle and Belle~II 
are weighted by $N_{\Upsilon(4S)}^{\rm B, B2} \times (\varepsilon_{i}^{\rm B, B2} \mathcal{B}_{i})$ 
and then combined. Assuming these uncertainties are independent and adding them in quadrature, the DE
uncertainties  are evaluated to be 2.7\% and 2.4\%  for the charged and neutral channels, respectively.

The statistical uncertainty on the simulation-based efficiency is at most 1.0\%.
The uncertainties on the branching fractions of
$\Lambda_{c}^{+}\to pK^{-}\pi^{+}$,
$\Lambda_{c}^{+}\to pK_{S}^{0}$,
$K_{S}^{0}\to\pi^{+}\pi^{-}$,
$\bar{\Xi}_{c}^{-}\to\bar{\Xi}^{+}\pi^{-}\pi^{-}$,
$\bar{\Xi}_{c}^{0}\to\bar{\Xi}^{+}\pi^{-}$,
$\bar{\Xi}_{c}^{0}\to\bar{\Lambda}K^{+}\pi^{-}$,
$\bar{\Xi}^{+}\to\bar{\Lambda}\pi^{+}$, and
$\bar{\Lambda}\to\bar{p}\pi^{+}$,
and on the ratio $R=\mathcal{B}(\bar{\Xi}_{c}^{-}\to\bar{p}K^{+}\pi^{-})/ \mathcal{B}(\bar{\Xi}_{c}^{-}\to\bar{\Xi}^{+}\pi^{-}\pi^{-})$,
are taken from Ref.~\cite{ParticleDataGroup:2024cfk}.
The corresponding systematic uncertainties are evaluated by varying each
quantity by $\pm1\,\sigma$ and taking the resulting deviation of the measured
branching fraction from its nominal value as the corresponding uncertainty.
For the two $\Lambda_{c}^{+}$ decay modes and the two
$\bar{\Xi}_{c}^{0}$ decay modes, the correlations between the corresponding
branching fractions are taken into account using the correlation coefficients
provided in Ref.~\cite{ParticleDataGroup:2024cfk}.
Correlation coefficients of 32\% and 74\% are used for the
$\Lambda_{c}^{+}$ and $\bar{\Xi}_{c}^{0}$ decay modes, respectively.
For the $\bar{\Xi}_{c}^{-}$ decay modes, the uncertainties on $R$ and
$\mathcal{B}(\bar{\Xi}_{c}^{-}\to\bar{\Xi}^{+}\pi^{-}\pi^{-})$
are propagated independently.
Given the large systematic uncertainties arising from the
branching fractions of $\bar{\Xi}_{c}^{-}$ and $\bar{\Xi}_{c}^{0}$  decays,
which may be reduced by future measurements, their contributions are quoted
separately as a third uncertainty. The resulting relative systematic uncertainties due to the uncertainties on the
branching fractions of $\bar{\Xi}_{c}^{-}$, $\bar{\Xi}_{c}^{0}$, and the other intermediate states are
$\ensuremath{{}_{-33\%}^{+82\%}}$, $\ensuremath{{}_{-17\%}^{+20\%}}$, and 3.2\%, respectively.
The uncertainties of $N_{\Upsilon(4S)}^{\rm B, B2}$ 
are 1.4\% for Belle and 1.5\% for Belle~II, 
and are combined into a total uncertainty weighted by $N_{\Upsilon(4S)}^{\rm B, B2} \times \sum_{i}(\varepsilon_{i}^{\rm B, B2} \mathcal{B}_{i})$.
The uncertainties of $f_{+-}$ and $f_{00}$ are 2.1\% and 1.7\%, respectively.
To evaluate the uncertainty associated with the BCS strategy, 
we adopt an alternative BCS procedure in which the candidate with the minimum 
$|M(\gamma \bar{\Xi}_c^{-,0}) - m(\bar{\Xi}_c^{\prime-,0})|$ is retained instead of the nominal criterion, 
where $m(\bar{\Xi}_c^{\prime-,0})$ denotes the known mass of the $\bar{\Xi}_c^{\prime-,0}$ baryon~\cite{ParticleDataGroup:2024cfk}. 
The resulting change in the measured branching fraction is taken as the corresponding systematic uncertainty. 
This uncertainty is estimated to be 2.5\% and 3.6\% for the charged and neutral channels, respectively.

The systematic uncertainties associated with the fit models arise from the signal PDF, the background PDFs, 
the choice of $M(\gamma \bar{\Xi}_c^{-,0})$ sideband regions, the modeling of the SCF components, 
the parameters $f_{\rm sig}$ and $f_{\rm scf}$, 
and the possible peaking background from non-resonant $\Lambda_{c}^{+} \pi^{\pm}$ combinations in the $\Sigma_c(2455)^{++,0}$ signal region.
The signal PDF is replaced by a triple-Gaussian function, and the resulting systematic uncertainty is negligible.
To evaluate the systematic uncertainty associated with the background PDFs, pseudo-experiments are generated 
using a second-order polynomial or an exponential function, with parameters taken from the corresponding alternative fits to data, 
while keeping the signal component unchanged.
Each pseudo-experiment  is fitted with the nominal background model, 
and the largest difference between the mean fitted signal yield and the corresponding input yield  is taken as systematic uncertainty.
To assess the uncertainty associated with the choice of
$M(\gamma \bar{\Xi}_c^{-,0})$ sideband regions,
the lower and upper sideband regions are shifted simultaneously toward lower and higher mass,
respectively, by 1, 2, 3, 4, and $5\,\mathrm{MeV}$, with their widths kept unchanged.
For the largest variation, the mass range between the lowest and highest sideband boundaries
corresponds to about $15\, \sigma$ of the
$M(\gamma \bar{\Xi}_c^{-,0})$ resolution.
The largest deviation of 
the fitted signal yield from the nominal result  is assigned as the systematic uncertainty. 
The SCF-related uncertainty is evaluated by interchanging the SCF PDFs derived from 
the $M(\gamma \bar{\Xi}_c^{-,0})$ signal and sideband regions,  with the resulting change in the 
fitted signal yield taken as the systematic uncertainty. 
To estimate the systematic uncertainty associated with $f_{\rm sig}$, 
we treat $f_{\rm sig}$ as a free parameter and repeat the fit.
The deviation from the nominal result is taken as the systematic uncertainty.
The uncertainty related to $f_{\rm scf}$ is covered by the variations of 
the sideband regions and the interchange of the SCF PDFs, which probe the uncertainty in the SCF modeling.
The possible peaking background from non-resonant $\Lambda_{c}^{+} \pi^{\pm}$ combinations
to the $\Delta E$ distribution is studied using events in the $M(\Lambda_c^{+}\pi^{\pm})$ sideband regions, 
and a simultaneous fit to the $\Delta E$ distributions in the $M(\gamma\bar{\Xi}_c^{ -,0})$ 
signal region, the $M(\gamma\bar{\Xi}_c^{ -,0})$ sideband regions, and the $M(\Lambda_c^{+}\pi^{\pm})$ sideband regions 
for Belle and Belle~II datasets is performed.
The resulting change in the fitted signal yield is smaller than 0.5\% for both the charged and neutral channels, and is therefore neglected.
The systematic uncertainties associated with the background PDFs, 
the choice of $M(\gamma \bar{\Xi}_c^{-,0})$  sideband regions, the SCF components,
and the value of $f_{\rm sig}$ 
are estimated to be 2.6\% (3.1\%), 2.4\% (1.6\%),  1.5\% (2.8\%), and 2.0\% (2.0\%) 
for charged channel (neutral), respectively.
All of these contributions are summed in quadrature to obtain the total systematic 
uncertainty related to the fit models.

\begin{table}[htbp]
	\centering
	\caption{\label{table2} Summary of fractional systematic uncertainties (\%) on the $\mathcal{B}(B^{+} \to \Sigma_{c}(2455)^{++} \bar{\Xi}_{c}^{\prime-})$ and $\mathcal{B}(B^{0} \to \Sigma_{c}(2455)^{0} \bar{\Xi}_{c}^{\prime0})$. The sources of all uncertainties are described in the text.}
	\vspace{0.2cm}
	\renewcommand{\arraystretch}{1.2}
	\begin{tabular}{lcccccccc}
		\hline\hline
		& $\sigma_{\rm DE}$ & $\sigma_{\rm eff}$ & $\sigma_{\BR_{i}}$ & $\sigma_{N_{\Upsilon(4S)}}$ & $\sigma_{f_{+-/00}}$ & $\sigma_{\rm BCS}$ & $\sigma_{\rm fit}$ & $\sigma_{\rm total}$ \\
		\hline
		$B^{+}$  & 2.7 & 1.0 & 3.2  & 1.0 & 2.1 & 2.5 &  4.3 & 7.0 \\
		$B^{0}$  & 2.4 & 1.0 & 3.2  & 1.0 & 1.7 & 3.6 &  4.9 & 7.6 \\
		\hline
		\hline
	\end{tabular}
\end{table}

In summary, we report the first observation of the decays
$B^{+} \to \Sigma_{c}(2455)^{++}\bar{\Xi}_{c}^{\prime-}$ and 
$B^{0} \to \Sigma_{c}(2455)^{0}\bar{\Xi}_{c}^{\prime0}$, 
using data samples corresponding to 
$771.6 \times 10^{6}$ and $520.6 \times 10^{6}$ $\Upsilon(4S)$ events 
collected by the Belle and Belle~II experiments, respectively. 
This represents the first observation of $B$-meson decays into a pair of charmed 
baryon-antibaryon states that belong to the same $SU(3)$ flavor sextet.
The branching fractions are measured to be 
$\mathcal{B}(B^{+} \to \Sigma_{c}(2455)^{++} \bar{\Xi}_{c}^{\prime-})	 =   (1.69 \pm 0.31 \pm 0.12^{+1.39}_{-0.56}) \times 10^{-3}$
and $\mathcal{B}(B^{0} \to \Sigma_{c}(2455)^{0} \bar{\Xi}_{c}^{\prime0}) = (1.28 \pm 0.32 \pm 0.10^{+0.26}_{-0.22})  \times 10^{-3}$,
where the uncertainties are statistical, systematic, and
from the absolute branching fractions of $\bar{\Xi}_{c}^{-}$ and $\bar{\Xi}_{c}^{0}$ 
decays, respectively.
The measured branching fractions are two to three times larger than those of 
$B^{+} \to \Sigma_{c}(2455)^{++}\,\bar{\Xi}_{c}^{-}$ and 
$B^{0} \to \Sigma_{c}(2455)^{0}\,\bar{\Xi}_{c}^{0}$~\cite{Belle:2025nup}, 
despite the smaller available phase space for these decays. 
This observation may indicate nontrivial dynamical effects in baryonic $B$ decays that enhance or suppress specific decay modes.


This work, based on data collected using the Belle II detector, which was built and commissioned prior to March 2019,
was supported by
Higher Education and Science Committee of the Republic of Armenia Grant No.~23LCG-1C011;
Australian Research Council and Research Grants
No.~DP200101792, 
No.~DP210101900, 
No.~DP210102831, 
No.~DE220100462, 
No.~LE210100098, 
and
No.~LE230100085; 
Austrian Federal Ministry of Education, Science and Research,
Austrian Science Fund (FWF) Grants
DOI:~10.55776/P34529,
DOI:~10.55776/J4731,
DOI:~10.55776/J4625,
DOI:~10.55776/M3153,
and
DOI:~10.55776/PAT1836324,
and
Horizon 2020 ERC Starting Grant No.~947006 ``InterLeptons'';
Natural Sciences and Engineering Research Council of Canada, Digital Research Alliance of Canada, and Canada Foundation for Innovation;
National Key R\&D Program of China under Contract No.~2024YFA1610503,
and
No.~2024YFA1610504
National Natural Science Foundation of China and Research Grants
No.~11575017,
No.~11761141009,
No.~11705209,
No.~11975076,
No.~12135005,
No.~12150004,
No.~12161141008,
No.~12405099,
No.~12475093,
and
No.~12175041,
China Postdoctoral Science Foundation (CPSF)
under Grant No.~2024M760485, and China
Postdoctoral Fellowship Program of CPSF under Grant
No.~GZC20240303,
and Shandong Provincial Natural Science Foundation Project~ZR2022JQ02;
the Czech Science Foundation Grant No. 22-18469S,  Regional funds of EU/MEYS: OPJAK
FORTE CZ.02.01.01/00/22\_008/0004632 
and
Charles University Grant Agency project No. 246122;
European Research Council, Seventh Framework PIEF-GA-2013-622527,
Horizon 2020 ERC-Advanced Grants No.~267104 and No.~884719,
Horizon 2020 ERC-Consolidator Grant No.~819127,
Horizon 2020 Marie Sklodowska-Curie Grant Agreement No.~700525 ``NIOBE''
and
No.~101026516,
and
Horizon Europe Marie Sklodowska-Curie Staff Exchange project JENNIFER3 Grant Agreement No.~101183137 (European grants);
L’Institut National de Physique Nucl\'eaire et de Physique des
Particules (IN2P3) du CNRS under Project Identification No.
CNRS-IN2P3-14-PP-033
and L’Agence Nationale de la Recherche (ANR) under Grant No. ANR-23-CE31-
0018 and ANR-25-CE31-1333 (France);
BMFTR, DFG, HGF, MPG, and AvH Foundation (Germany);
Department of Atomic Energy under Project Identification No.~RTI 4002,
Department of Science and Technology,
and
UPES SEED funding programs
No.~UPES/R\&D-SEED-INFRA/17052023/01 and
No.~UPES/R\&D-SOE/20062022/06 (India);
Israel Science Foundation Grant No.~2476/17,
U.S.-Israel Binational Science Foundation Grant No.~2016113, and
Israel Ministry of Science Grant No.~3-16543;
Istituto Nazionale di Fisica Nucleare and the Research Grants BELLE2,
and
the ICSC – Centro Nazionale di Ricerca in High Performance Computing, Big Data and Quantum Computing, funded by European Union – NextGenerationEU;
Japan Society for the Promotion of Science, Grant-in-Aid for Scientific Research Grants
No.~16H03993,
No.~16H06492,
No.~16K05323,
No.~17H01133,
No.~17H05405,
No.~18K03621,
No.~18H03710,
No.~18H05226,
No.~19H00682, 
No.~20H05850,
No.~20H05858,
No.~22H00144,
No.~22K14056,
No.~22K21347,
No.~23H05433,
No.~26220706,
No.~26400255,
and
No.~26H02056,
and
the Ministry of Education, Culture, Sports, Science, and Technology (MEXT) of Japan;  
National Research Foundation (NRF) of Korea Grants
No.~2021R1-F1A-1064008,
No.~2022R1-A2C-1003993,
No.~RS-2018-NR031074,
No.~RS-2021-NR060129,
No.~RS-2024-00354342,
No.~RS-2025-02219521,
No.~RS-2026-25471491,
No.~RS-2026-25480677,
and
No.~RS-2026-25486791,
Radiation Science Research Institute,
Foreign Large-Size Research Facility Application Supporting project,
the Global Science Experimental Data Hub Center, the Korea Institute of Science and
Technology Information (K26L1M2C3)
and
KREONET/GLORIAD;
Universiti Malaya RU grant, Akademi Sains Malaysia, and Ministry of Education Malaysia;
Frontiers of Science Program Contracts
No.~FOINS-296,
No.~CB-221329,
No.~CB-236394,
No.~CB-254409,
and
No.~CB-180023, and SEP-CINVESTAV Research Grant No.~237 (Mexico);
the Polish Ministry of Science and Higher Education and the National Science Center;
the Ministry of Science and Higher Education of the Russian Federation
and
the HSE University Basic Research Program, Moscow;
University of Tabuk Research Grants
No.~S-0256-1438 and No.~S-0280-1439 (Saudi Arabia);
Slovenian Research Agency and Research Grants
No.~J1-50010
and
No.~P1-0135;
Ikerbasque, Basque Foundation for Science,
State Agency for Research of the Spanish Ministry of Science and Innovation through Grant No. PID2022-136510NB-C33, Spain,
Agencia Estatal de Investigacion, Spain
Grant No.~RYC2020-029875-I
and
Generalitat Valenciana, Spain
Grant No.~CIDEGENT/2018/020;
The Knut and Alice Wallenberg Foundation (Sweden), Contracts No.~2021.0174, No.~2021.0299, and No.~2023.0315;
National Science and Technology Council,
and
Ministry of Education (Taiwan);
Thailand Center of Excellence in Physics;
TUBITAK ULAKBIM (Turkey);
National Research Foundation of Ukraine, Project No.~2020.02/0257,
and
Ministry of Education and Science of Ukraine;
the U.S. National Science Foundation and Research Grants
No.~PHY-1913789 
and
No.~PHY-2111604, 
and the U.S. Department of Energy and Research Awards
No.~DE-AC06-76RLO1830, 
No.~DE-SC0007983, 
No.~DE-SC0009824, 
No.~DE-SC0009973, 
No.~DE-SC0010007, 
No.~DE-SC0010073, 
No.~DE-SC0010118, 
No.~DE-SC0010504, 
No.~DE-SC0011784, 
No.~DE-SC0012704, 
No.~DE-SC0019230, 
No.~DE-SC0021616, 
No.~DE-SC0022350, 
No.~DE-SC0023470; 
and
the Vietnam Academy of Science and Technology (VAST) under Grant
No.~DL0000.05/26-27.

These acknowledgements are not to be interpreted as an endorsement of any statement made
by any of our institutes, funding agencies, governments, or their representatives.

We thank the SuperKEKB team for delivering high-luminosity collisions;
the KEK cryogenics group for the efficient operation of the detector solenoid magnet and IBBelle on site;
the KEK Computer Research Center for on-site computing support; the NII for SINET6 network support;
and the raw-data centers hosted by BNL, DESY, GridKa, IN2P3, INFN, 
and the University of Victoria.

\section*{Data Availability}
\ifdefvoid{\hepdata}{}{Numerical data corresponding to the results presented are available as HEPData.}
The full Belle II data are not publicly available. The collaboration will consider requests for access to the data that support this article.

\appendix
\section{$M(\Lambda_{c}^{+}\pi^{\pm})$ versus $M(\gamma\bar{\Xi}_{c}^{-,0})$ distributions }~\label{App_A}

\begin{figure*}[htbp]
	\begin{center}
		\hspace{-1.5cm}
		\includegraphics[width=6.8cm]{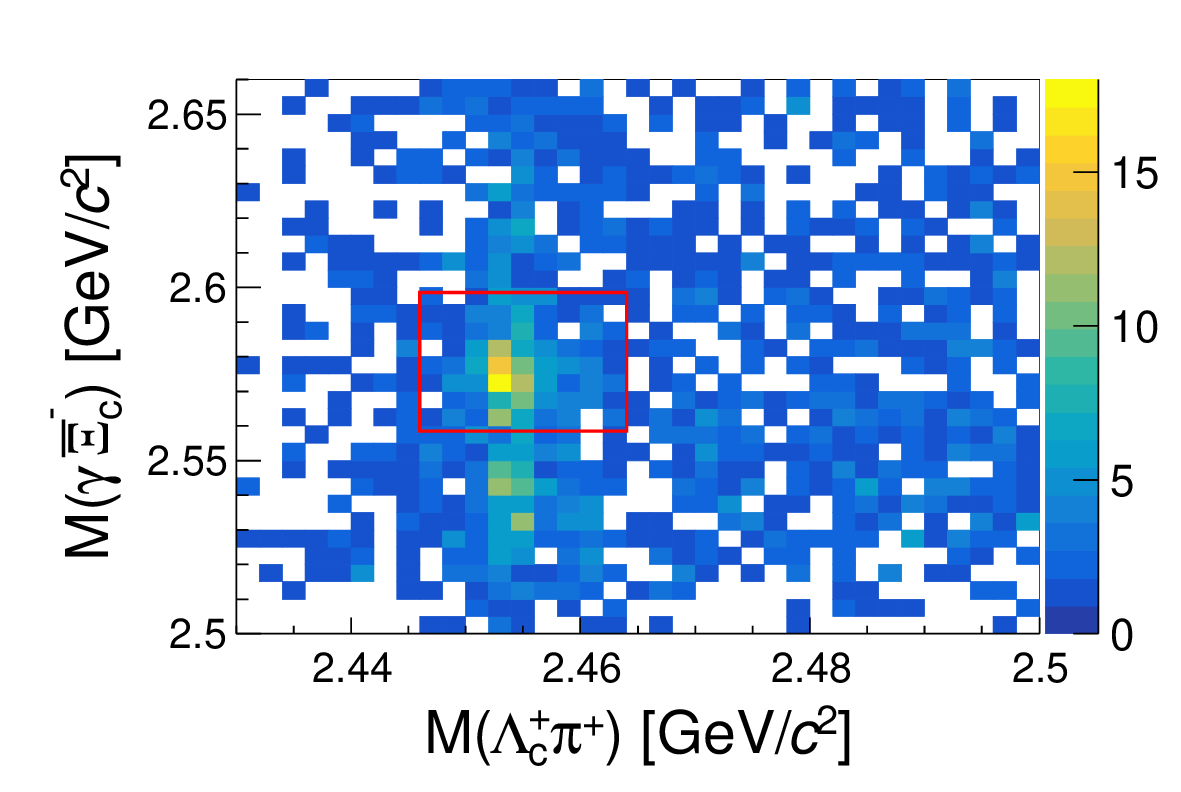}
		\includegraphics[width=6.8cm]{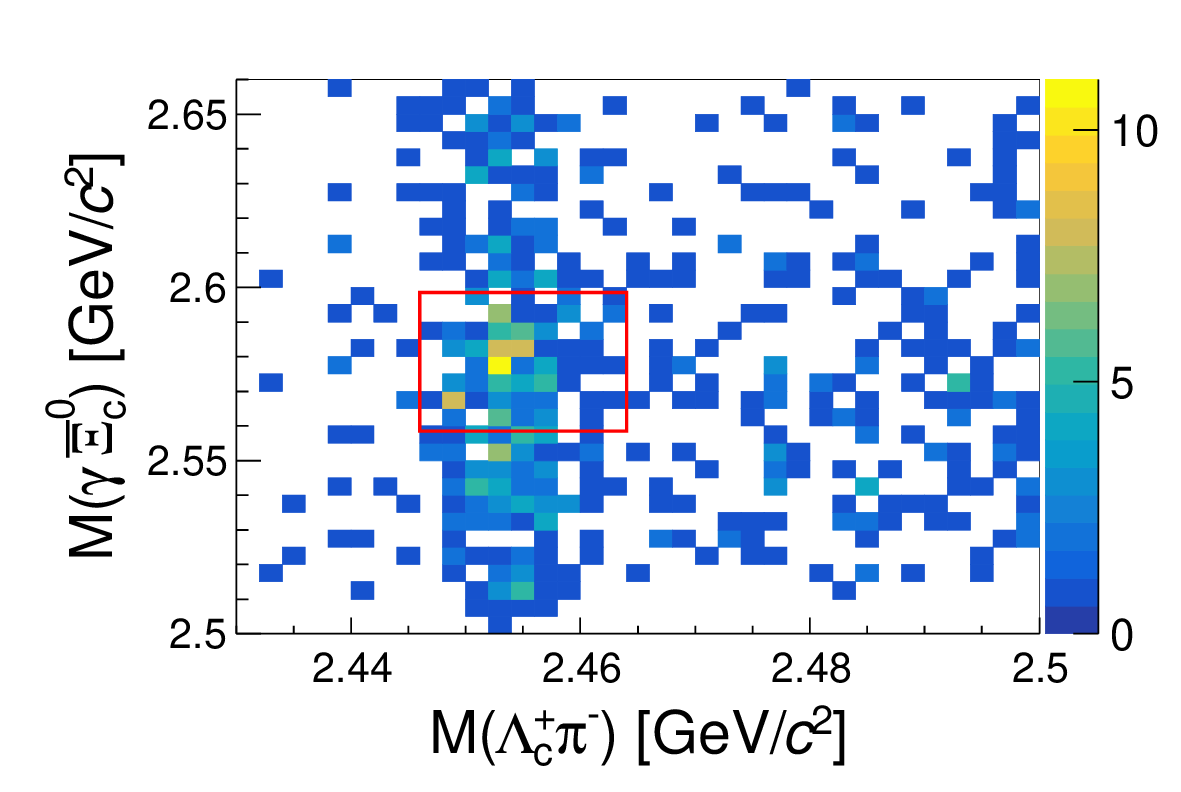}
		\put(-290,130){$\int \lum dt$ = 711\,fb$^{-1}$ (Belle) + 492\,fb$^{-1}$ (Belle~II)}
		
		\hspace{-1.5cm}
		\includegraphics[width=6.8cm]{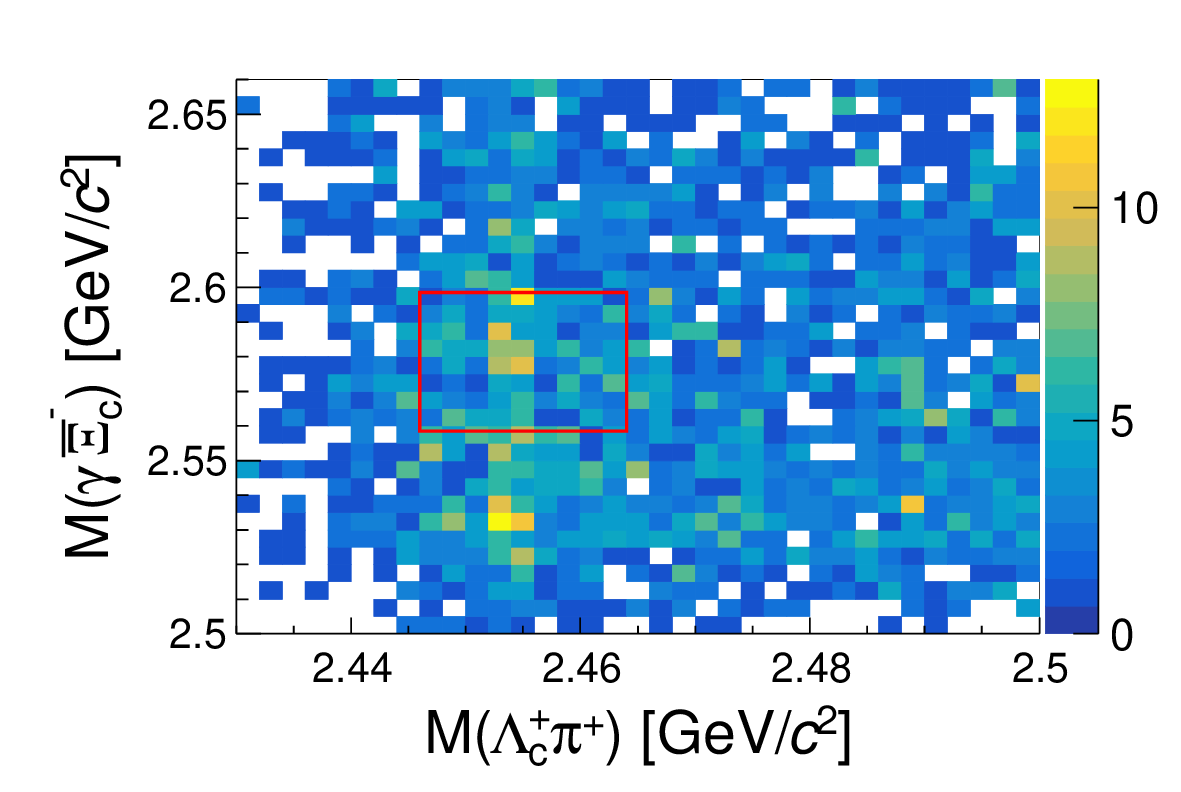}
		\includegraphics[width=6.8cm]{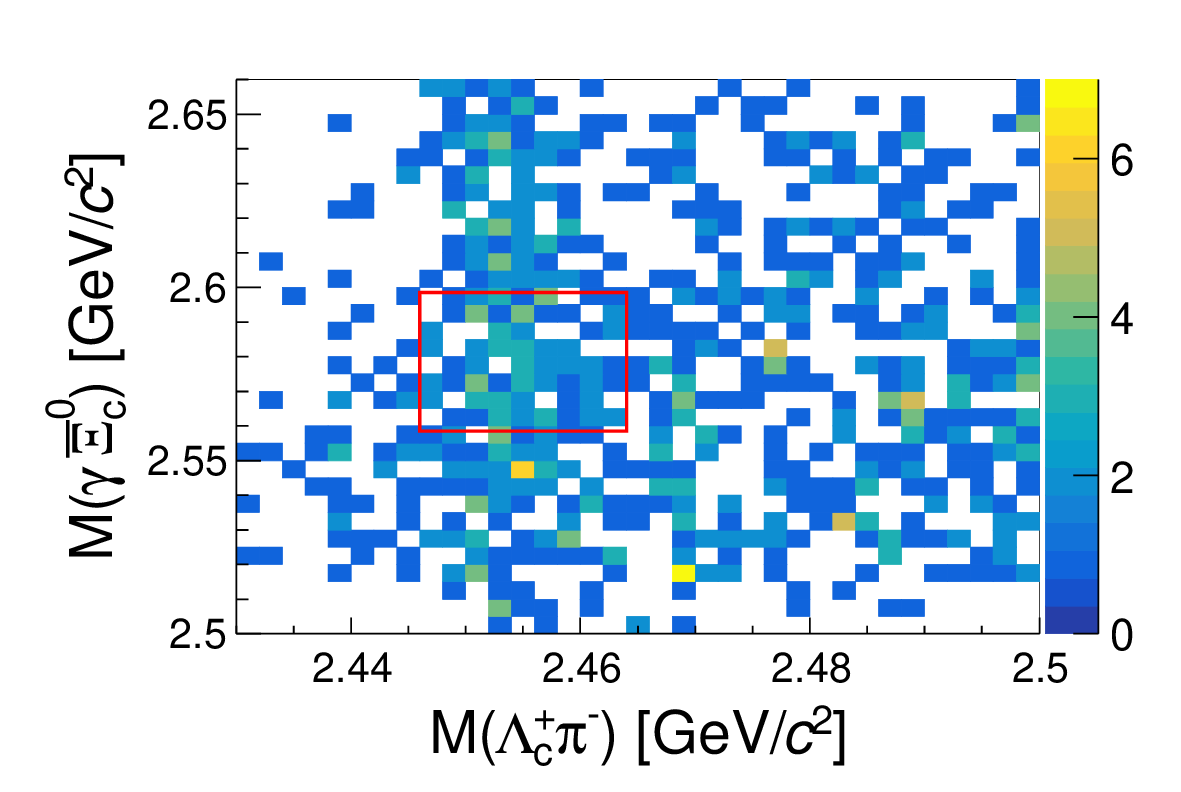}
   \caption{Scatter plots of $M(\Lambda_{c}^{+}\pi^{\pm})$ versus $M(\gamma\bar{\Xi}_{c}^{-,0})$ for reconstructed 
	$B^{+}\to\Sigma_{c}(2455)^{++}\bar{\Xi}_{c}^{\prime-}$ (left) and 
	$B^{0}\to\Sigma_{c}(2455)^{0}\bar{\Xi}_{c}^{\prime0}$ (right) candidates 
	in the  signal (top) and sideband (bottom) regions of $\Delta E$ from the combined 
	Belle and Belle~II data samples, before application of the BCS. 
	The red boxes indicate the signal regions for the 
	$\Sigma_c(2455)^{++,0}$ and $\bar{\Xi}_c^{\prime-,0}$ candidates.} \label{fig3}
	\end{center}
\end{figure*}

\begin{figure*}[htbp]
	\begin{center}
		\hspace{-0.7cm}
		\includegraphics[width=7cm]{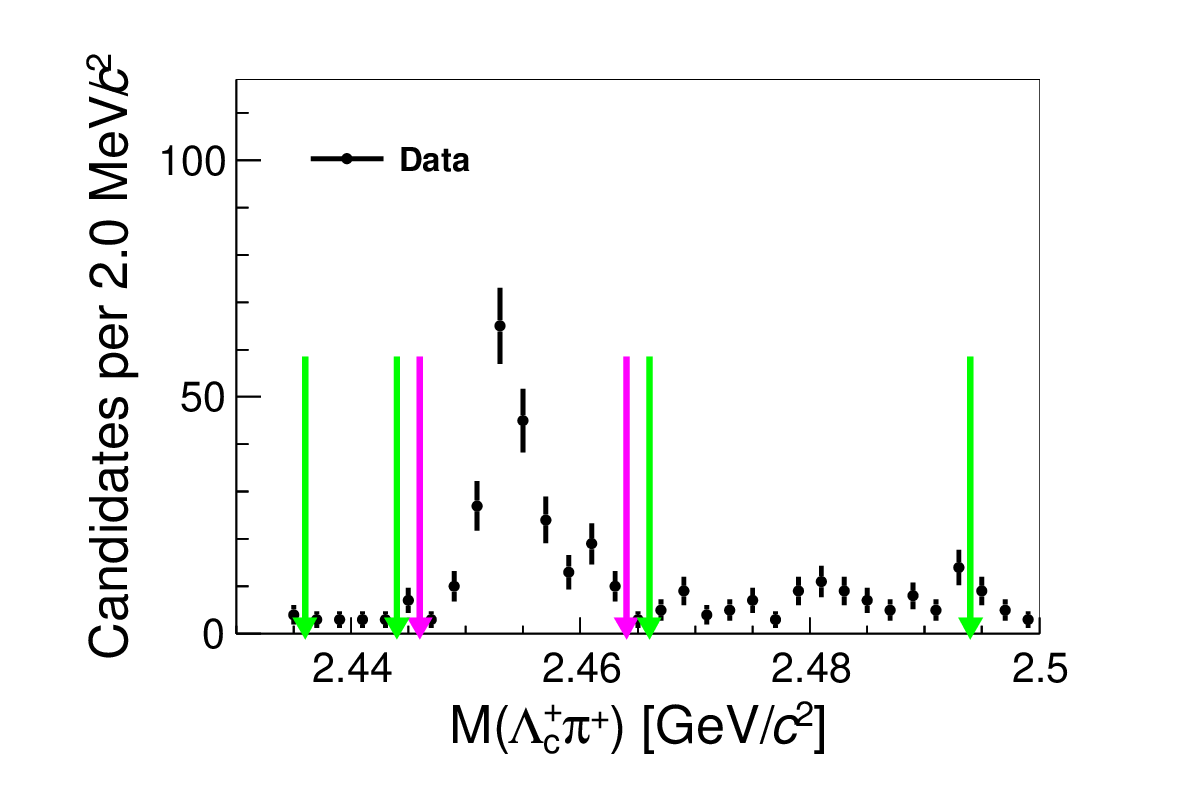}\hspace{-0.3cm}
		\includegraphics[width=7cm]{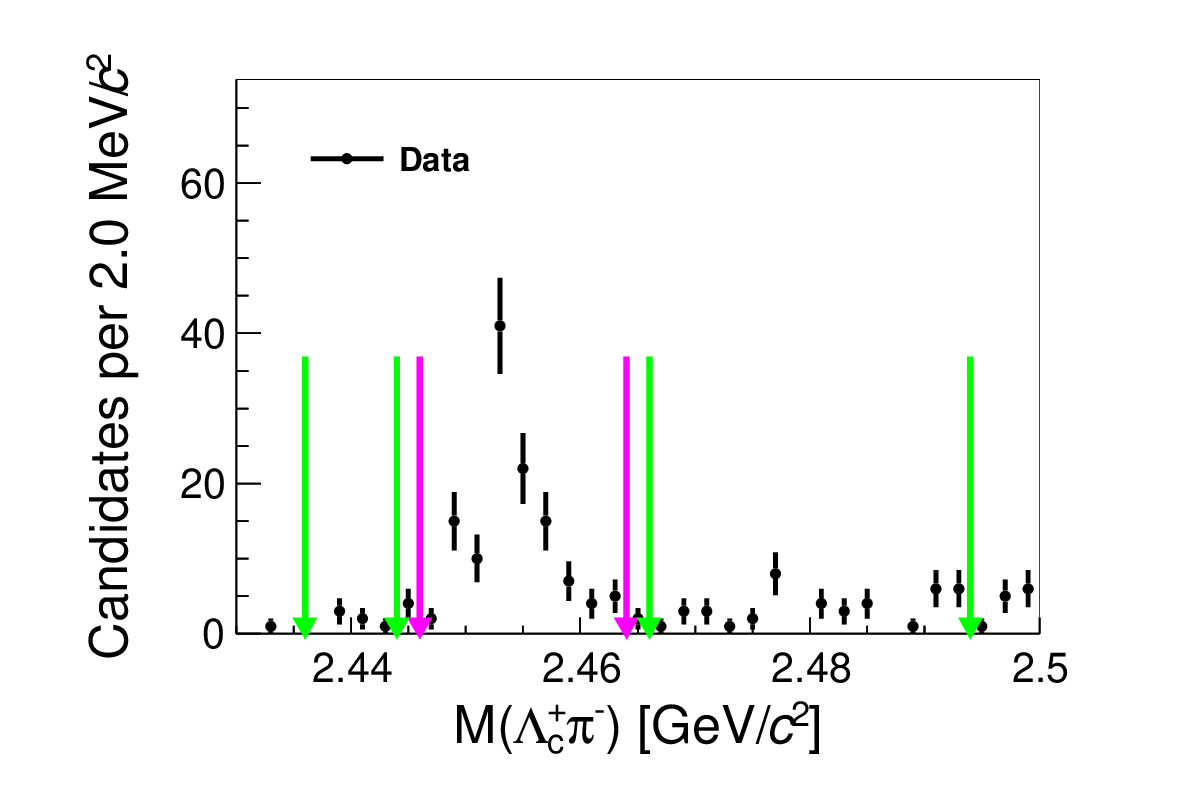}
		\put(-290,140){$\int \lum dt$ = 711\,fb$^{-1}$ (Belle) + 492\,fb$^{-1}$ (Belle~II)}
		
		\hspace{-0.7cm}
		\includegraphics[width=7cm]{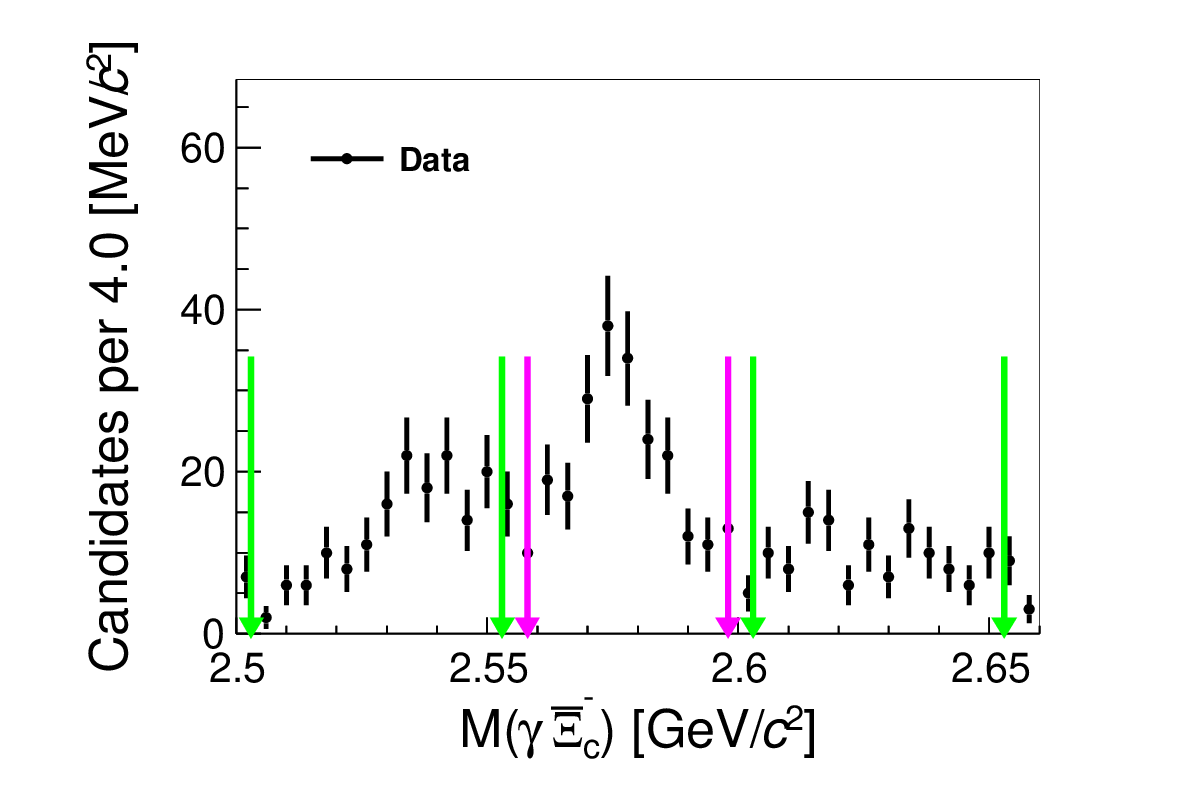}\hspace{-0.3cm}
		\includegraphics[width=7cm]{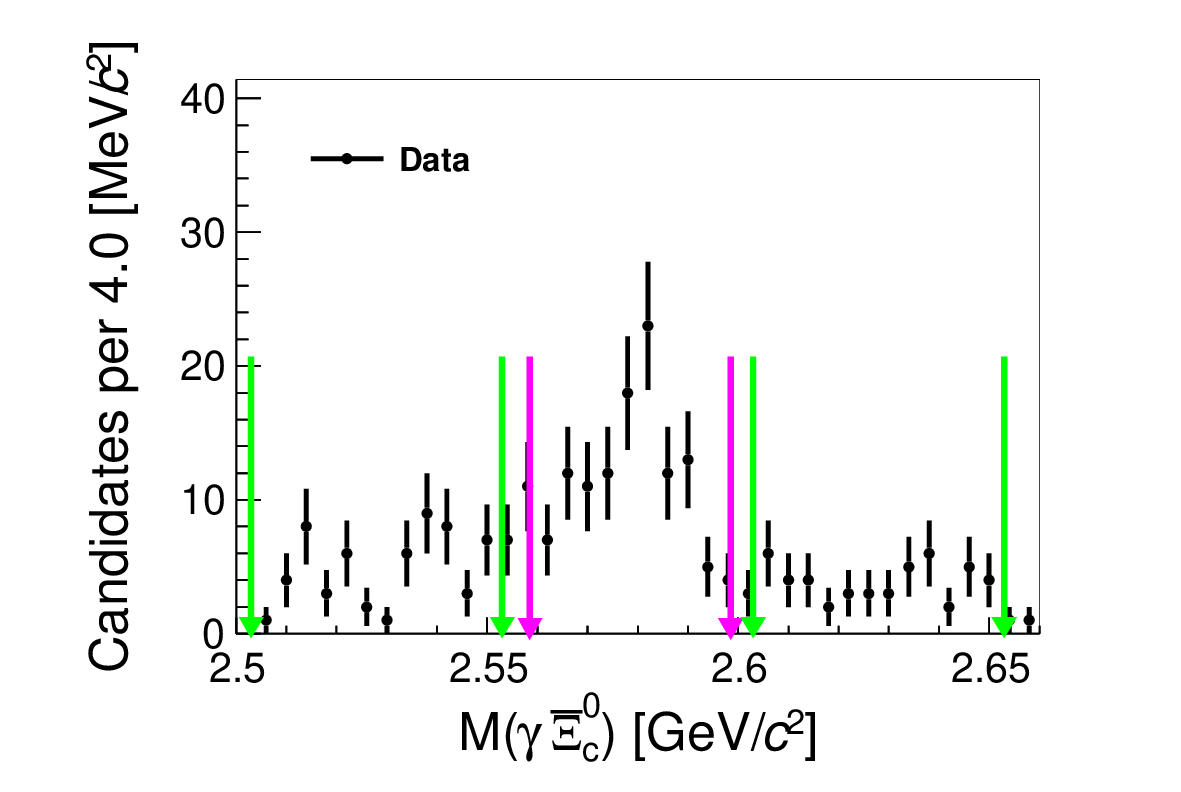}
      \caption{One-dimensional projections onto (top)  $M(\Lambda_{c}^{+}\pi^{\pm})$ and (bottom) 
	  $M(\gamma\bar{\Xi}_{c}^{-,0})$ for reconstructed 
 	  $B^{+}\to\Sigma_{c}(2455)^{++}\bar{\Xi}_{c}^{\prime-}$ (left) and 
	  $B^{0}\to\Sigma_{c}(2455)^{0}\bar{\Xi}_{c}^{\prime0}$ (right) candidates 
	  in the  signal region of $\Delta E$ from combined Belle and Belle~II data samples, 
	  before application of the BCS. Points with error bars represent the data. 
	 The magenta arrows indicate the signal regions defined for the 
	 $\Sigma_{c}(2455)^{++,0}$ and $\bar{\Xi}_{c}^{\prime-,0}$ candidates, 
	 while the green arrows indicate the corresponding sideband regions.}\label{fig4}
	\end{center}
\end{figure*}

Figure~\ref{fig3} shows the scatter plots of 
$M(\Lambda_{c}^{+}\pi^{\pm})$ versus 
$M(\gamma\bar{\Xi}_{c}^{-,0})$ for reconstructed 
$B^{+}\to\Sigma_{c}(2455)^{++}\bar{\Xi}_{c}^{\prime-}$ and 
$B^{0}\to\Sigma_{c}(2455)^{0}\bar{\Xi}_{c}^{\prime0}$ candidates 
in the signal and sideband regions of $\Delta E$ from the combined 
Belle and Belle~II data samples, before application of the BCS. 
The  signal region of $\Delta E$ is defined as 
$|\Delta E|<0.012~\mathrm{GeV}$ (approximately  $2.5\, \sigma$), while the sideband regions are defined as 
$0.030<|\Delta E|<0.054~\mathrm{GeV}$, with a total width twice that of the signal region. 
The corresponding one-dimensional projections onto 
$M(\Lambda_{c}^{+}\pi^{\pm})$ and $M(\gamma\bar{\Xi}_{c}^{-,0})$ 
for candidates in the signal region of $\Delta E$ are shown in 
Fig.~\ref{fig4}. Clear signals corresponding to 
$\Sigma_c(2455)^{++,0}$ and $\bar{\Xi}_c^{\prime-,0}$ are observed.

\section{Reconstruction efficiencies for the individual reconstruction modes}\label{App_B}
Tables~\ref{table3} and~\ref{table4} list the reconstruction efficiencies and the products of the reconstruction 
efficiencies and the corresponding branching fractions of all intermediate states for the individual reconstruction modes of the
$B^{+}\to\Sigma_{c}(2455)^{++}\bar{\Xi}_{c}^{\prime-}$ and $B^{0}\to\Sigma_{c}(2455)^{0}\bar{\Xi}_{c}^{\prime0}$ decays, respectively.

\begin{table*}[htbp]
	\centering
	\caption{\label{table3}
		Reconstruction efficiencies ($\varepsilon_i$) and the products of the reconstruction efficiencies and 
		the branching fractions of all intermediate states ($\varepsilon_i\mathcal{B}_{i}$) for the individual reconstruction modes of the
		$B^{+} \to \Sigma_{c}(2455)^{++}(\to \Lambda_{c}^{+} \pi^{+}) \bar{\Xi}_{c}^{\prime-}(\to \gamma \bar{\Xi}_{c}^{-})$ decay.
		The branching fractions of all intermediate states are taken from Ref.~\cite{ParticleDataGroup:2024cfk}.
		The branching fraction $\BR(\bar{\Xi}_{c}^{-}\to\bar{p}K^{+}\pi^{-})$ is inferred from the ratio
		$\BR(\bar{\Xi}_{c}^{-}\to\bar{p}K^{+}\pi^{-})/\BR(\bar{\Xi}_{c}^{-}\to\bar{\Xi}^{+}\pi^{-}\pi^{-})$
		and the absolute branching fraction $\BR(\bar{\Xi}_{c}^{-}\to\bar{\Xi}^{+}\pi^{-}\pi^{-})$.}
	\renewcommand{\arraystretch}{1.35}
	\begin{tabular}{lcccc}
		\hline\hline
		& \multicolumn{2}{c}{Belle}
		& \multicolumn{2}{c}{Belle~II} \\
		\cline{2-3}\cline{4-5}
		Reconstruction mode
		& $\varepsilon_i$ (\%)
		& $\varepsilon_i\mathcal{B}_i$ ($10^{-5}$)
		& $\varepsilon_i$ (\%)
		& $\varepsilon_i\mathcal{B}_i$ ($10^{-5}$) \\
		\hline
		$\Lambda_c^{+}\to pK^{-}\pi^{+}$,\quad
		$\bar{\Xi}_c^{-}\to\bar{\Xi}^{+}\pi^{-}\pi^{-}$
		& 0.70 & 0.83 & 1.04 & 1.23 \\
		
		$\Lambda_c^{+}\to pK^{-}\pi^{+}$,\quad
		$\bar{\Xi}_c^{-}\to\bar{p}K^{+}\pi^{-}$
		& 3.06 & 1.19 & 3.98 & 1.54 \\
		
		$\Lambda_c^{+}\to pK_S^{0}$,\quad
		$\bar{\Xi}_c^{-}\to\bar{\Xi}^{+}\pi^{-}\pi^{-}$
		& 0.82 & 0.17 & 1.17 & 0.25 \\
		
		$\Lambda_c^{+}\to pK_S^{0}$,\quad
		$\bar{\Xi}_c^{-}\to\bar{p}K^{+}\pi^{-}$
		& 3.43 & 0.24 & 4.40 & 0.30 \\
		\hline
		$\sum_i\varepsilon_i\mathcal{B}_i$
		& \multicolumn{2}{c}{$2.43\times10^{-5}$}
		& \multicolumn{2}{c}{$3.32\times10^{-5}$} \\
		\hline\hline
	\end{tabular}
\end{table*}

\begin{table*}[htbp]
	\centering
	\caption{\label{table4}
	Reconstruction efficiencies ($\varepsilon_i$) and the products of the reconstruction efficiencies and 
	the branching fractions of all intermediate states ($\varepsilon_i\mathcal{B}_{i}$) for the individual reconstruction modes of the
	$B^{0} \to \Sigma_{c}(2455)^{0}(\to \Lambda_{c}^{+} \pi^{-}) \bar{\Xi}_{c}^{\prime0}(\to \gamma \bar{\Xi}_{c}^{0})$ decay.
	The branching fractions of all intermediate states are taken from Ref.~\cite{ParticleDataGroup:2024cfk}.}
	\renewcommand{\arraystretch}{1.35}
	\begin{tabular}{lcccc}
		\hline\hline
		& \multicolumn{2}{c}{Belle}
		& \multicolumn{2}{c}{Belle~II} \\
		\cline{2-3}\cline{4-5}
		Reconstruction mode
		& $\varepsilon_i$ (\%)
		& $\varepsilon_i\mathcal{B}_i$ ($10^{-5}$)
		& $\varepsilon_i$ (\%)
		& $\varepsilon_i\mathcal{B}_i$ ($10^{-5}$) \\
		\hline
		$\Lambda_{c}^{+}\to pK^{-}\pi^{+}$,\quad
		$\bar{\Xi}_{c}^{0}\to\bar{\Xi}^{+}\pi^{-}$
		& 1.04 & 0.61 & 1.46 & 0.85 \\
		
		$\Lambda_{c}^{+}\to pK^{-}\pi^{+}$,\quad
		$\bar{\Xi}_{c}^{0}\to\bar{\Lambda}K^{+}\pi^{-}$
		& 1.13 & 0.67 & 2.05 & 1.22 \\
		
		$\Lambda_{c}^{+}\to pK_{S}^{0}$,\quad
		$\bar{\Xi}_{c}^{0}\to\bar{\Xi}^{+}\pi^{-}$
		& 1.18 & 0.12 & 1.67 & 0.17 \\
		
		$\Lambda_{c}^{+}\to pK_{S}^{0}$,\quad
		$\bar{\Xi}_{c}^{0}\to\bar{\Lambda}K^{+}\pi^{-}$
		& 1.28 & 0.13 & 2.33 & 0.25 \\
		\hline
		$\sum_i\varepsilon_i\mathcal{B}_i$
		& \multicolumn{2}{c}{$1.53\times10^{-5}$}
		& \multicolumn{2}{c}{$2.49\times10^{-5}$} \\
		\hline\hline
	\end{tabular}
\end{table*}

\end{document}